\documentstyle[manuscript,aps,psfig]{revtex}
\tighten
\begin{document}
\draft
\title{Dilepton Enhancement by Thermal Pion Annihilation 
       in the CERES Experiment}
\author{H.-J. Schulze and D. Blaschke}
\address{MPG-AG ``Theoretische Vielteilchenphysik'', Universit\"at Rostock,
         D-18051 Rostock, Germany}
\maketitle
\date{}
\begin{abstract}
 
We compare the recent CERES data on dielectron production in
200 GeV/u S+Au collisions with the theoretical predictions
due to pion annihilation
in a thermal hadronization and a string fragmentation scenario.
Both models yield similar results for the dilepton mass spectrum.
A satisfactory description of the experimental spectrum 
requires an in-medium reduction of the rho-mass and a freeze-out 
temperature of about 150 MeV in the thermal model. 
We emphasize and discuss the influence of experimental 
acceptance and resolution corrections.

\end{abstract}
\vskip1.5cm

%\section{Introduction}
The recent results of the CERES collaboration on the production of 
electron pairs in S+Au collisions at $200\;\rm GeV/u$ 
\cite{ceres,ceresi} allow for the first time a {\em quantitative}
comparison with theoretical models for this type of ultrarelativistic
heavy ion collisions.
The CERES collaboration reports a five-fold enhancement of the dilepton
spectrum in the mass range from the two-pion threshold up to 
$1.5 \;\rm GeV$, 
compared to the expectation based on an independent
superposition of nucleon-nucleon collisions.
Their interpretation of this enhancement is being due to the production
of dileptons from pion-pion annihilation in the hot and dense medium that is 
supposedly formed during the collision.  
The shape of the observed spectrum then hints to a modification of the
pion annihilation form factor in this medium.
There have been a couple of quantitative analyses of the results 
based on relativistic transport models \cite{cass,lkb,crw}.
These works are able to describe the data reasonably well
using a suitable medium-modified form factor.

On the other hand, thermal models have been used successfully to describe 
observables like hadron abundances \cite{bms} and hadron spectra \cite{heinz}
in ultrarelativistic heavy ion collisions.
While giving relatively good fits to the data,
they have the advantage of simplicity and 
and a small number of adjustable parameters,
the most prominent one being a ``freeze-out'' temperature $T_f$.

%It is therefore of interest to investigate the prediction of such a thermal
%hadronization 
In this paper we would like to analyze the dilepton production 
within a similarly simple thermal hadronization model, 
the standard ``benchmark'' scenario
based on the Bjorken model of a boost-invariant 
longitudinal expansion \cite{bjorken},
including a quark phase (u+d+g), a mixed phase, 
and a hadronic pion gas phase.
It is characterized by three temperatures, 
the formation temperature $T_i$ of the quark phase, 
%which can be fixed by the parameters of the collision, 
the critical temperature $T_c$ for the hadronization in a mixed
phase of quarks and pions,
%with a "canonical" value of $T_c=160$ MeV, 
and the freeze-out temperature $T_f$.
%which is an adjustable parameter of the model.
This scenario has often been employed to predict the dilepton production rate 
in ultrarelativistic heavy ion collisions
\cite{thmod,distr}, until recently however without a chance of  
quantitative experimental comparison.

A first test of the model will be the comparison with a purely 
hadronic scenario based on the pion distribution functions 
as obtained from a string-fragmentation event generator 
(in this case VENUS) \cite{hjo}.
Both models should yield similar results for the (unmodified)
dilepton mass spectrum, because at present collision energies
the quark annihilation contribution in the thermal model is negligible 
compared to the pion annihilation part \cite{thmod}.
This supposition is indeed confirmed by our results.
The models therefore provide a reliable basis upon which more sophisticated
features like medium-dependent properties of pion \cite{pimod} 
or rho-meson \cite{rhomod,rhoconst,brown},
finite pion \cite{pimu1,pimu2,pimu3} or quark \cite{qmu} chemical potentials, 
transverse expansion \cite{qmu,trexp}, 
or more exotic possibilities \cite{weldon}, can be studied.
We will then use the thermal model in order to test the influence of 
a medium modification of the pion form factor. 
With the simple model of an unmodified form factor in the hadronic phase
and a dropping rho-mass in the mixed phase we are able to achieve a very 
satisfactory fit of the dilepton spectrum when the thermal history of the 
hot matter is described by a set of temperatures 
$(T_i,T_c,T_f) = (250,160,150)\;\rm MeV$.

\begin{figure}
\begin{picture}(0,500)
 \put(-20,-100){\psfig{figure=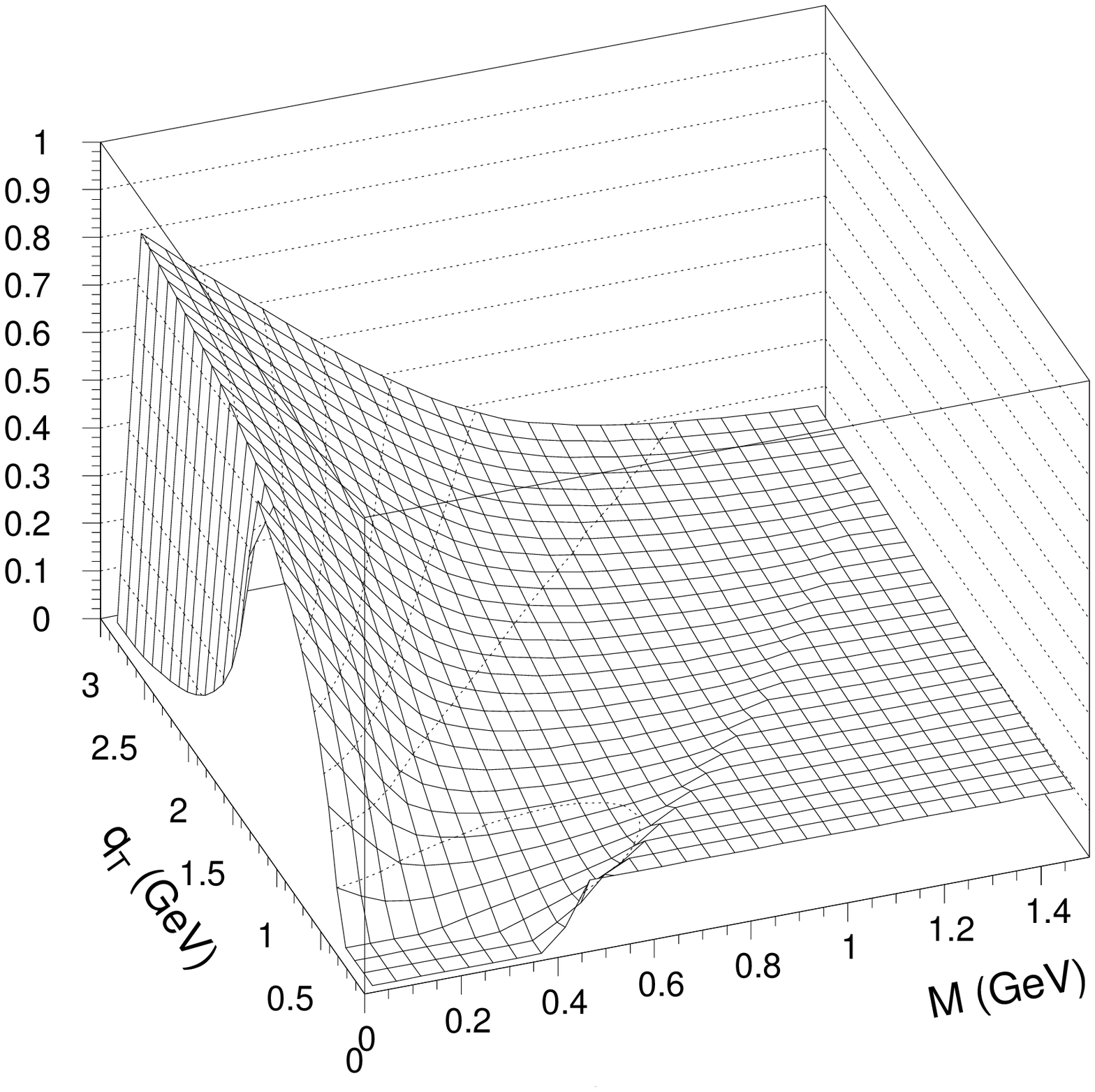,width=18cm,angle=0}}
%% \put(-40,-400){\epsfxsize=10cm\epsffile{gapav.ps}}
%% \put(-40,-650){\special{epsf=paris.eps scale=1.} }
\end{picture}
\caption{
  The acceptance function for a virtual photon with mass $M$ and 
  transverse momentum $q_T$ decaying into an 
  electron pair subject to the kinematical restrictions 
  Eq.~(\protect\ref{e:cuts}). } 
\label{f:accep}
\end{figure}

Before a detailed description of the models, 
we start however with a discussion of the 
general ``kinematical'' conditions of the experiment.
Obviously, in any comparison with a theoretical model, 
the same kinematical cuts and detector resolutions as in the actual
experiment have to be applied.
In this case, they are fairly severe:
A dilepton is only detected, if both the electron and the positron momenta
satisfy the kinematical restrictions on pseudorapidity, transverse momentum,
and opening angle,
\begin{mathletters}
\begin{eqnarray}
 &&  2.1 < \eta_e < 2.65 \ ,
\\
 &&  p_T^{(e)} > 0.2\;{\rm GeV} \ , 
\\
 &&  \Theta_{ee} > 35\;{\rm mrad} = 2^\circ \ . 
\label{e:cut3}
\end{eqnarray}
\label{e:cuts}
\end{mathletters}
These conditions translate into an acceptance function $A(M,q_T)$
depending on the mass $M$ and transverse momentum $q_T$ of the 
virtual photon that decays into the lepton pair.
A first guess can be made by neglecting the transverse momentum.
The acceptance 
for an isotropic decay of the virtual photon in its rest frame 
(as appropriate for a thermal production model)
is then simply given by 
\begin{equation}
  A(M,0) \approx \theta(M-0.4\;{\rm GeV}) {\log[\cosh(0.275)]\over 0.275} \ .
\end{equation}
The numerical value of the second factor is 0.136, and the observed 
spectrum is therefore reduced by nearly an order of magnitude.
The full acceptance function can be computed numerically and is shown in 
Fig.~\ref{f:accep}.
(In fact it turns out that for two-body decays 
the restriction on the opening angle
Eq.~(\ref{e:cut3}) has only an effect for extremely 
low masses $M < 50\;\rm MeV$, as displayed in the figure). 
It can be seen that there is a strong variation with transverse momentum, 
in particular for low masses that we are interested in.
The notion of an acceptance function $A(M)$ 
depending only on the dilepton mass $M$, 
%as employed in Ref.~\cite{crw}, 
is therefore a strong approximation.
For a proper treatment the theoretical spectrum $dN_{ee}/d^4q$ has 
to be multiplied with the acceptance function $A(M,q_T)$, 
before the mass spectrum $dN_{ee}/ dM dy$
can be computed by integration over $q_T$. 
Before a comparison with the experimental data, the theoretical spectrum 
has still to be folded with a Gaussian representing 
the finite mass resolution $\delta M/ M$ of the detector, 
which is approximately given by the formula \cite{ceresi}:
\begin{equation}
  {\delta M\over M} =
  \left\{ \begin{array}{ll}
            0.08                  & \quad ,\quad M < 0.5\; {\rm GeV} \\
            0.03+0.1M/{\rm GeV}  & \quad ,\quad M > 0.5\; {\rm GeV} \ .
          \end{array} \right.
\end{equation}
Finally, for a careful comparison it is important to note that 
the CERES result represents an average over an ensemble of events with 
mean charged particle multiplicity 
$\langle dN_{\rm ch}/dy \rangle \approx 125$ and RMS value
$\sqrt{ \langle (dN_{\rm ch}/dy)^2 \rangle} \approx 135$ \cite{ceresi}. 

We come now to the detailed description of both scenarios of
pion annihilation in order to compute the dilepton spectrum $dN_{ee}/d^4q$.
Starting with a short review of the thermal model, 
the production rate of electron pairs 
with four-momentum $q$ from the relativistic quark or pion gas
at temperature $T$ is
\begin{equation}
  { dN_{ee} \over d^4x d^4q } = { M^2 a \sigma(M) \over 4\pi(2\pi)^4 }
  F(M,E,T)
\label{e:dndxdq}
\end{equation}
with $M^2=q^2$, $a=\sqrt{1-4m^2/M^2}$, $(m=m_q,m_\pi)$, 
and the elementary cross sections in the quark (Q) 
and hadronic (H) phases,
\begin{mathletters}
\begin{eqnarray}
  a\sigma(M) = {4\pi\alpha^2\over3 M^2} \widetilde\sigma(M) 
  \quad,\quad 
  && \widetilde\sigma_Q(M) =
  {20\over 3} \left( 1+{2m_q^2\over M^2} \right)  
  \left(1+{2m_e^2\over M^2}\right) \sqrt{1-{4m_e^2\over M^2}} \ ,
\\
  && \widetilde\sigma_H(M) =
  |F_\pi(M)|^2 \left( 1-{4m_\pi^2\over M^2} \right)
  \left(1+{2m_e^2\over M^2}\right) \sqrt{1-{4m_e^2\over M^2}} 
\end{eqnarray}
\end{mathletters}
with the pion form factor
\begin{equation}
  |F_\pi(M)|^2 = { C m_\rho^4 + m_\rho^2\Gamma_\rho^2
  \over 
  (M^2-m_\rho^2)^2 + m_\rho^2 \Gamma_\rho^2 }
\label{e:fpi}
\end{equation}
and $C=1.3$, $m_\rho=760\;\rm MeV$, $\Gamma_\rho=135\;\rm MeV$. 
(These parameters fit the experimental data \cite{piff} better than the 
standard choice $C=1$, $m_\rho=770\;\rm MeV$, $\Gamma_\rho=160\;\rm MeV$).

The function $F(M,E,T)$ in Eq.~(\ref{e:dndxdq}) 
is given by a phase-space integral over
the Fermi (Bose) distributions of the
reacting quarks (pions) \cite{distr,pimu2}:
\begin{mathletters}
\begin{eqnarray}
  F_Q(M,E,T) &=& {1\over \beta \bar q}{1\over \exp{(\beta E)}+1}
  \log\left[ { \cosh{\beta \omega_{\rm max}} + \cosh{\beta\mu_q} \over
               \cosh{\beta \omega_{\rm min}} + \cosh{\beta\mu_q} } \right] \ ,
\\
  F_H(M,E,T) &=& 
  {1\over \beta \bar q}{1\over \exp{(\beta E - 2\beta\mu_\pi)}-1}
  \log{\left[ { \cosh{\beta(\omega_{\rm max}-\mu_\pi)} - 1 \over
                \cosh{\beta(\omega_{\rm min}-\mu_\pi)} -1  } \right] } \ .
\end{eqnarray}
\end{mathletters}
Here $E=q\cdot u(x)$ is the dilepton energy, and
$\bar q = \sqrt{E^2-M^2}$ is the three-momentum
component of $q$ in the production fluid element moving 
with four-velocity $u$.
$\omega_{\rm max,min} = (E\pm a \bar q)/2$
is the maximal (minimal) quark or pion energy in this frame. 
The formula includes finite chemical potentials for the 
quarks \cite{qmu} and the pions \cite{pimu1,pimu2}, 
although we restrict ourselves in the following to the case
$\mu_q=\mu_\pi=0$.
The integration of the rate over space-time then gives the 
momentum spectrum of lepton pairs:
\begin{eqnarray}
  {dN_{ee}\over d^4q} &=& 
  \pi R^2 \int_0^\infty d\tau \tau \int_{-\infty}^{\infty} d\eta \;
  {dN_{ee}\over d^4x d^4q}
\\
  &=& {\alpha^2 \over 16 \pi^5 } 
      { (\pi R^2 \tau_c T_c^3 )^2 \over R^2 M_T^6 } \times
\nonumber
\\
  &&
      \Bigg\{ \;\widetilde\sigma_Q(M) 
      \left[ \int_{M_T/T_i}^{M_T/T_c} dz z^5 G_Q(M,M_T,T={M_T/ z})
             + {r-1\over 6} z_c^6 G_Q(M,M_T,T_c)  \right] +
\nonumber   
\\
  && \phantom{+\;} 
      \widetilde\sigma_H(M) 
      \left[ r^2 \int_{M_T/T_c}^{M_T/T_f} dz z^5 G_H(M,M_T,T=M_T/z)
             + r {r-1\over 6} z_c^6 G_H(M,M_T,T_c)  \right] \;\Bigg\} \ ,
\label{e:dndq}
\end{eqnarray}
where $r=37/3$ is the ratio of degrees of freedom in the 
quark and hadronic phases, and
$G$ represents the integral over space-time rapidity:
\begin{equation}
  G(M,M_T,T) = \int_{-\infty}^\infty d\eta \, F(M,E,T) 
               \Bigr|_{ E(\eta)= M_T\cosh(\eta-y) } \ .
\end{equation}
It sums the dileptons produced in different slices of 
space-time rapidity $\eta$ at given proper-time $\tau$ (and temperature $T$).
The energy $E$ of the dilepton in these local production frames is 
related by a Lorentz-boost to the energy in the lab system, $M_T \cosh y$.
The normalization of the spectrum can be linked to the square of the
pion rapidity density by employing the relation 
$\pi R^2 \tau T^3 = \kappa\, dN_\pi/dy$, with $\kappa \approx 0.22$
in the quark phase.
The final result is 
\begin{equation}
  {dN_{ee}\over d^4q} = 
  {\alpha^2\over 16 \pi^5}
  {(\kappa\, dN_\pi/dy)^2 \over R^2} {1\over M_T^6} 
  \Biggl\{ \ldots \Biggr\}
  \approx { 7\times 10^{-8} \;{\rm GeV}^2 \over M_T^6 }  
  \Biggl\{ \ldots \Biggr\} \ ,
\end{equation}
where the expression in curly brackets is the one given 
in Eq.~(\ref{e:dndq}).
In the last step we have inserted the appropriate numerical values 
$R\approx R_S \approx 18\,{\rm GeV}^{-1}$, and 
$\langle dN_\pi/dy\rangle_{\rm RMS} \approx 200$ \cite{ceresi}.

\begin{figure}
\begin{picture}(0,550)
 \put(-40,-100){\psfig{figure=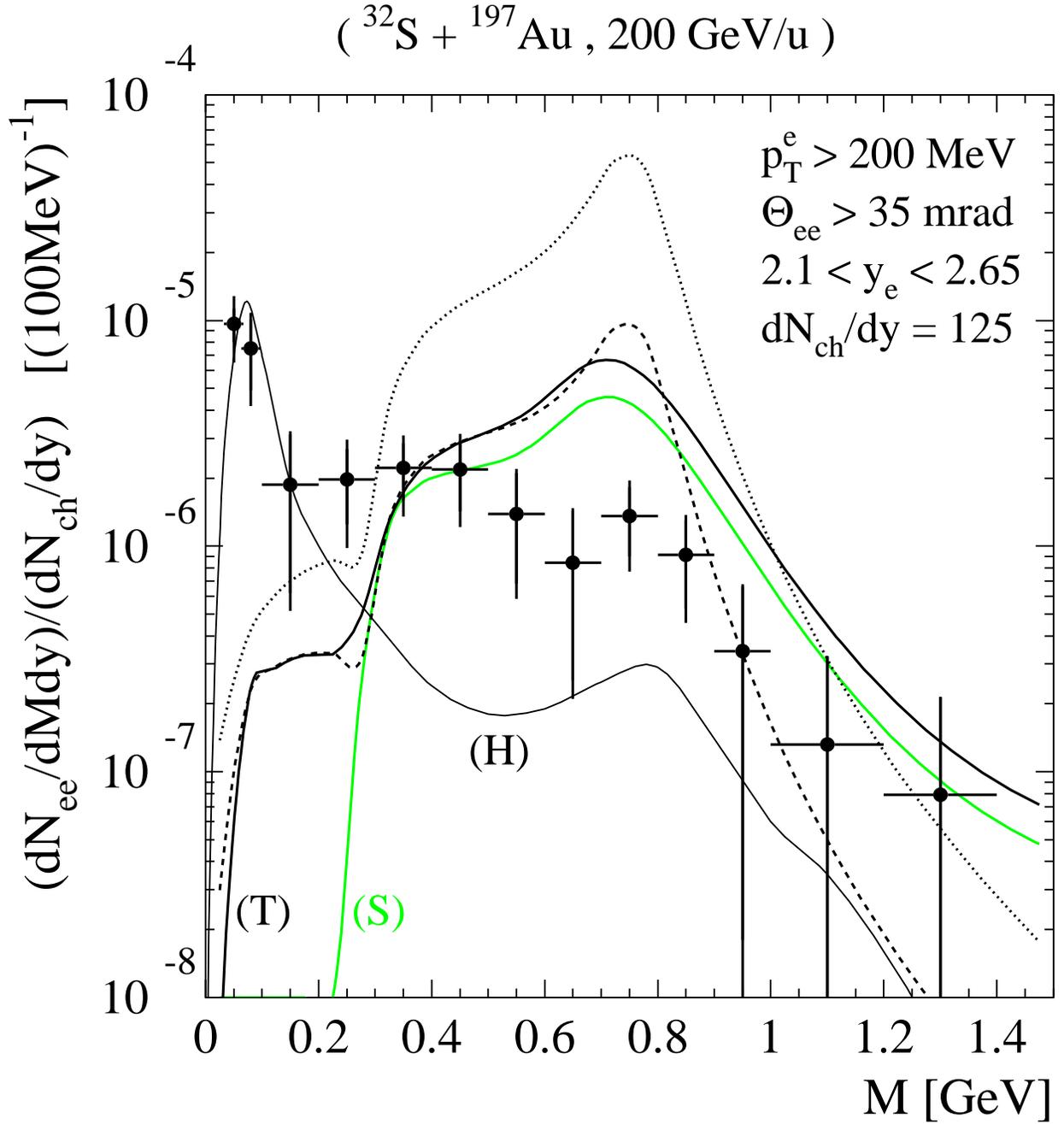,width=20cm,angle=0}}
\end{picture}
\caption{
  The dilepton mass spectrum. The full lines represent the theoretical 
  predictions in the thermal (T) and the string model (S) scenario after 
  acceptance and resolution corrections.
  For the thermal scenario, we plot also the results before resolution
  correction (dashed line) and before acceptance correction (dotted line). 
  The data are taken from Ref.~\protect\cite{ceres}. 
  (Statistical and systematical errors added quadratically). 
  The curve denoted (H) shows the total hadron decay contribution 
  as given in that reference. }
\label{f:dilep}
\end{figure}

The spectrum is now completely determined by specifying the three
temperatures $T_i$, $T_c$, and $T_f$, and
we use here for the moment a standard set of $T_i=250\;\rm MeV$, 
$T_c=160\;\rm MeV$, and $T_f=120 \;\rm MeV$, that we will discuss later.
The comparison with the experimental mass spectrum can then be done by 
multiplying the theoretical spectrum with the acceptance function, 
integrating over the transverse momentum, and folding with the 
mass resolution of the detector, as discussed above.
The result is shown in Fig.~\ref{f:dilep}, where we display the 
spectrum before any correction, after acceptance correction, and the 
final result.
The comparison of the three curves demonstrates the importance
of the corrections.
In particular the slope of the mass spectrum beyond the rho-peak 
seems to be entirely determined by the finite mass resolution.
Before a further discussion of the figure, we turn however
to the second, completely hadronic scenario of dilepton 
production, namely by pion annihilation in a string fragmentation
model.

We start by repeating the expression for the dilepton 
spectrum found in Ref.~\cite{hjo}:
\begin{equation}
 { dN_{ee} \over d^4 q} = 
 { 16 \alpha^2 \over 27{\langle p_T \rangle}^4{\langle m_T \rangle}} \;
 \left[ \int_0^\infty d\tau \; 
        {dN_\pi\over dy}(y,\tau)\epsilon_\pi(y,\tau) \right] 
 {1 \over \cosh^4(y)}\; 
 { \widetilde\sigma_H(M) \exp(-M_T/a) \over M M_T } 
\label{e:pian}
\end{equation}
with $d^4q = M dM dy d^2q_T$.
This result was obtained by assuming that throughout the reaction 
the pions are produced with a momentum distribution
$dN_\pi/dy dp_T^2 \sim \delta(y-\eta) \exp(-m_T/a)$.
The parameters $\langle p_T \rangle$ and $\langle m_T \rangle$
are the average transverse momentum and transverse mass of the pions.
For further information on the assumptions of the model, 
and details of the calculation, we refer to Ref.~\cite{hjo}.

The dynamical content of Eq.~(\ref{e:pian}) is the
one-dimensional integral of the temporal development of the
product of particle density $dN_\pi/dy$ and energy density
$\epsilon_\pi = \langle m_T\rangle/(\tau \pi R^2)\; dN_\pi/dy$
in a given slice of rapidity,
\begin{equation}
  I = \int_0^\infty d\tau \; 
  {dN_\pi\over dy}(y,\tau)\,\epsilon_\pi(y,\tau) \ . 
\label{e:piint}
\end{equation}
This integral contains the dependence on energy and mass number of 
the colliding nuclei.
%It converges due to the transverse expansion of the interaction region.
In Ref.~\cite{hjo} we used the results of the VENUS event generator 
\cite{venrep} on particle and energy densities for central 
{O+Au} collisions at $200 \;\rm GeV/u$ \cite{venus} 
and found an upper limit of $40\;\rm GeV^3$ for this integral.
Unfortunately, new results of VENUS on these observables for the case of a
{S+Au} reaction are not available.
We therefore employ here a simple scaling assumption and multiply the 
{O+Au} result with a factor
\begin{equation}
  \left( {(dN_\pi/dy)_S \over (dN_\pi/dy)_O} \right)^2
  \times {R_O^2 \over R_S^2}
  \approx 1.1 \ ,
\label{e:scl}
\end{equation}
with $(dN_\pi/dy)_O\approx 150$ and $(dN_\pi/dy)_S\approx 200$ 
appropriate for the two cases.
Considering the accuracy of the whole model,
this procedure should certainly be adequate.
Further we use in Eq.~(\ref{e:pian})
a kinematical factor $\cosh^4{(y)}\approx 2.1$,
since the center of the CERES rapidity coverage lies about 0.6 units 
away from the nucleon-nucleon cms rapidity (3.0).
The resulting dilepton spectrum, 
with an average transverse momentum of pions 
$\langle p_T \rangle \approx 0.4\;\rm GeV$ \cite{pionpt}, 
and after acceptance and resolution corrections, 
is also shown in Fig.~\ref{f:dilep}.

It is immediately noted that although the two scenarios considered here, 
thermal and string fragmentation, rest on completely different physical
assumptions and operate on different time scales,
their predictions, at least for the mass spectrum of dileptons, are nearly
undistinguishable 
(apart from the quark phase contribution in the thermal model at 
very low masses, which is however completely negligible in the present 
type of reaction).
This shows that on the one hand the dilepton mass spectrum is 
unfortunately rather insensitive to the underlying hadron (pion) dynamics, 
on the other hand it allows us to use now the thermal model
with some confidence in order to adjust parameters and 
test assumptions about medium modifications. 

Concerning the influence of the choice of temperatures $T_i$, $T_c$, $T_f$,
we find that a variation of $T_i$ has practically no importance, since the 
quark part of the spectrum is negligible anyway.
The value of the critical temperature $T_c=160\;\rm MeV$ appears generally
accepted, and we fix it as a ``canonical" value.
A variation of $T_f$ within a relevant range 
$100 \ldots 150\;\rm MeV$ changes the spectrum by about a factor two,
but preserves its overall shape.
A low value of $T_f=120\;\rm MeV$, as used in Fig.~\ref{f:dilep}, 
is a more ``traditional'' choice \cite{thmod,distr},  
whereas rather large values are supported by 
recent fits to hadron abundances 
in the present experiment ($T_f \approx 160-170$ MeV, \cite{bms}) 
or to slopes of hadron spectra ($T_f\approx 150$ MeV, \cite{heinz}). 

%Still, we stress that we do not perform a
%``fit'' of the normalization of the
%theoretical spectra, like in Ref.~\cite{crw}.
We find then that with $T_f=120$ (150) MeV and using 
the free pion form factor the theoretical predictions exceed
the CERES data by about a factor four (three) in the rho-mass region.
In comparison, in Refs.~\cite{lkb} and \cite{crw} that are based on more 
sophisticated transport models, the corresponding
enhancement factors are approximately three and two, respectively.
%This excess could be attributed to the neglect of the transverse expansion
%in the thermal model,
%(it has been shown \cite{trexp} that the transverse expansion can reduce 
%the hadronic part of the spectrum by even an order of magnitude),
%respectively to the fact that the integral Eq.~(\ref{e:piint}) is
%evaluated at the radial center of the reaction.
We can now conclude that a sizable finite chemical potential of the 
pions \cite{pimu1,pimu2} can probably be excluded by our result, 
since this would 
again {\em enhance} the dilepton spectrum by nearly an order of
magnitude, as shown in Ref.~\cite{pimu2}.

A possibility of reducing the theoretical expectation 
is of course the in-medium modification of the 
elementary cross sections and form factors, 
in particular for the rho meson \cite{medmod1,medmod}. 
The pion is thought to be rather stable against medium effects \cite{pimod}, 
and we do not consider a modification here.
Unfortunately, the numerous models on the modification of 
the rho meson \cite{rhomod,rhoconst,brown} 
do not even agree on the simplest qualitative aspects of the variation
of, say, the rho-mass $m_\rho$ and width $\Gamma_\rho$ in the medium.

%A couple of models is inspired by... 
%In Ref.~\cite{}... 
We do not intend to introduce a further model here. 
Instead we will use two simple parametrizations of the temperature 
dependences of $m_\rho$ and $\Gamma_\rho$ 
in the form factor Eq.~(\ref{e:fpi})
together with the thermal scenario described above 
in order to assess the principle effects on the dilepton spectrum.

The first parametrization is the one employed in Ref.~\cite{medmod1}, 
namely an increase of the width with temperature, 
while the mass is kept constant:
%\begin{equation}
%  |F_\pi(M)|^2 = { C\, m_\rho^4 +m_\rho^2\Gamma_\rho^2(T) \over 
%  (M^2-m_\rho^2)^2 + m_\rho^2 \Gamma_\rho^2(T) }
%  \quad,\quad
%  \Gamma_\rho(T) = {\Gamma_\rho \over 1-T^2/T_c^2 }
%\end{equation}
$ \Gamma_\rho(T) = {\Gamma_\rho / (1-T^2/T_c^2) }$.
This means that in the mixed phase the pion-pion cross section is 
reduced to the Born term, 
while in the hadronic phase it is still substantially suppressed.
This is probably unrealistic, and
the results should therefore be considered as a lower bound on what
a medium modification of the form factor can achieve. 

The second parametrization is based on the assumption that the  
vector meson mass does not change dramatically in a hot hadronic medium
and is during the hadronic phase given by its vacuum value \cite{rhoconst}. 
However, in the mixed phase at the critical temperature 
for the deconfinement transition $T_c=160\;\rm MeV$, 
the mass might be lowered as conjectured 
by a number of authors \cite{brown}. 
Here we simply assume a mass of 
$m_\rho(T_c) = m_\rho/2 \approx 380\;\rm MeV$.
Although at present there is no firm theoretical foundation for such an
assumption, we consider it as a guess which allows to 
reproduce the two-peak structure of the experimental spectrum.

\begin{figure}
\begin{picture}(0,550)
 \put(-40,-100){\psfig{figure=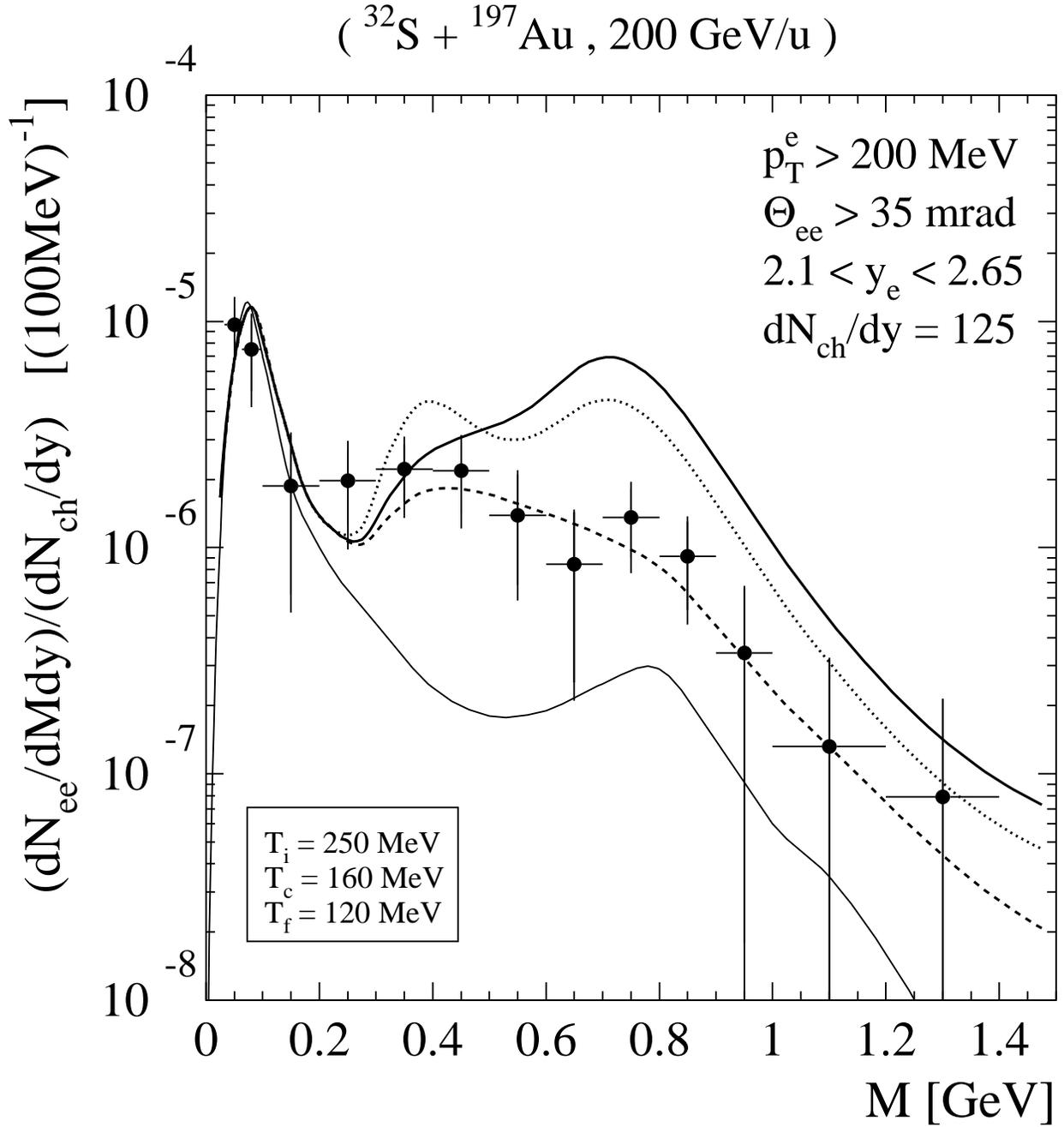,width=20cm,angle=0}}
\end{picture}
\caption{
  The dilepton mass spectrum with an in-medium modification of the
  pion form factor and $T_f=120\;\rm MeV$. 
  The dashed curve shows the result with a 
  temperature dependent width, and the dotted curve with a temperature 
  dependent rho mass, as explained in the text. 
  For comparison the full curve based on the vacuum
  pion form factor is also shown again. 
  The hadron decay contribution (thin line) has been added to our pion
  annihilation results. }
\label{f:tdep}
\end{figure}

\begin{figure}
\begin{picture}(0,550)
 \put(-40,-100){\psfig{figure=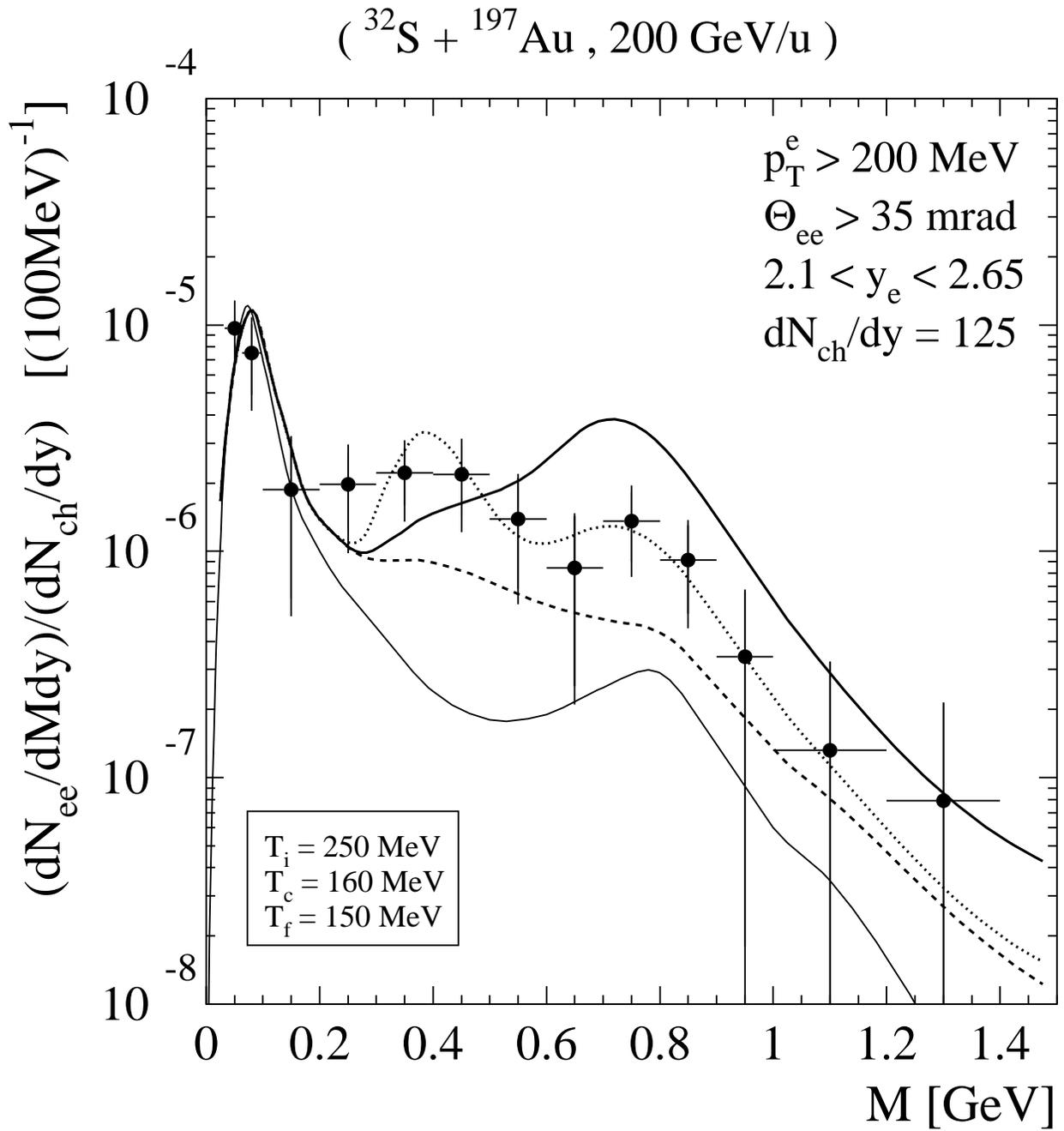,width=20cm,angle=0}}
\end{picture}
\caption{
  Same as Fig.~\protect\ref{f:tdep}, 
  but with a temperature $T_f=150\;\rm MeV$. }
\label{f:tdep2}
\end{figure}

We display the results in Figs.~\ref{f:tdep},\ref{f:tdep2}.
For a better comparison with the data, we have added here the 
part of the dilepton spectrum due to hadron decays, as given in 
Ref.~\cite{ceres}, to our pion annihilation results.
Indeed the spectrum below 200 MeV is well explained
by the Dalitz decays of $\pi_0$, $\eta$, and $\omega$ mesons.
We observe then that the varying-width scenario that we consider as a lower
bound, is in good agreement with the overall normalization of the data 
using $T_f=120\;\rm MeV$, 
whereas the dropping-mass model at $T_f=150\;\rm MeV$ gives a rather good fit 
to the shape of the spectrum, 
in fact of the same quality as obtained with sophisticated transport models
and more realistic medium modifications \cite{cass,lkb,crw}.
This shows that the larger freeze-out temperature of 
150 MeV is better suited to fit the data,
in agreement with the fairly large freeze-out temperatures
deduced from the fit of hadron abundances \cite{bms} in the same
experiment. 
A suitable in-medium modification of the pion form factor 
%perhaps together with an overall reduction factor of about 
%two to three due to transverse expansion, 
is thus indeed able to account for the observed spectrum.

This statement is valid with the exception of the data point at 
$M=0.25\;\rm MeV$, i.e., below the two-pion threshold, that neither in 
our nor in previous models \cite{cass,lkb,crw} can be reproduced in a 
satisfactory manner. 
This hints to the possibility of a 
non-negligible modification of the pion dispersion relation in the medium, 
but also the Dalitz decay of the $\eta$ meson is still important in
this region and should be reconsidered.

We conclude that pion annihilation can indeed account for the 
observed excess of dileptons,
as already advocated by the CERES collaboration \cite{ceres}.
We have presented a detailed adaptation of both a simple Bjorken scenario 
for the thermal dilepton production, 
as well as of a purely hadronic scenario due to 
string fragmentation, to the experimental situation.
These models have the advantage of transparency and simplicity, 
but can not be expected to give perfect fits to the data.
Nevertheless, we find that both models give surprisingly similar results
(which means that the possible existence of a quark-gluon plasma can neither
 be affirmed nor rejected from the present data),
and can account for the experimental spectrum by assuming 
a dropping rho-mass in the medium. 
The justification of such an assumption remains a theoretical challenge.
We have here merely provided a guess that allows to reproduce the data.
A freeze-out temperature of about 150 MeV is supported by the 
experimental results.
The correct implementation of acceptance and resolution corrections
has been seen to be extremely important.

In particular with a large freeze-out temperature of about 150 MeV 
nearly all dileptons then stem from pion annihilation in the rather
long-lived mixed phase 
(of duration $(r-1)\tau_c \approx 24\;{\rm fm}/c$ 
 in the one-dimensional expansion scenario,
 and with the parameters chosen here).
This puts forward the question of physical reality of such a phase.
First principle, i.e., lattice gauge calculations
do not yet provide an unanimous answer to this problem \cite{latt}.
It is possible that a first order transition does not occur at all, 
or that the release of latent heat is fairly small.
In any case there exists a relatively narrow transition region 
between the phases, in which hadronic correlations persist.
The simple model of a first order phase transition between ideal
quark and hadron gases is certainly an oversimplification, and should
be considered a phenomenological approach to mimic the features of 
this real situation.
The same remark applies of course to other features of the computation,
such as the treatment of collision dynamics or medium modification of 
hadron parameters, as discussed before. 

We would finally like to add a speculative estimate of the 
expected dilepton spectrum in central 
Pb+Au (or Pb+Pb) collisions \cite{pb}.
In both models presented here, the dominant source of change
is a scaling factor ${(dN_\pi/dy)^2 / R^2}$,
cf.~Eq.~(\ref{e:scl}).
For example, with $(dN_\pi/dy)_{\rm Pb}\approx 700$ 
we would obtain a rescaling of the curves presented in this paper 
by about a factor $12/3.4 \approx 3.5$.
Of course, this scaling factor applies only with identical acceptance cuts
as in the present experiment, so that in practice a completely 
new analysis suited to the experimental conditions will have to be done.

\bigskip
We would like to thank A.~Drees and Th.~Ullrich for discussions and detailed
information on the setup of the CERES experiment.
We thank G.~R\"opke for a critical reading of the manuscript.
D.B. acknowledges enlightening discussions with K. Redlich.

%------------------------------------------------------------------------------

%-----------------------------------------------------------------------------

%-----------------------------------------------------------------------------
\end{document}